\documentclass[12pt]{article}
\usepackage{graphicx}

\begin{document}

\begin{center}
{\bf  \LARGE Predicting Baseball Home Run Records Using Exponential Frequency Distributions}

\vskip 1.5 cm
Daniel J. Kelley, Jonas R. Mureika, Jeffrey A. Phillips \\
\vskip .2cm
{\small \it Department of Physics, Loyola Marymount University\\
1 LMU Drive, MS-8227, Los Angeles, CA  90045 \\
Contact E-mail: jphillips@lmu.edu}
\end{center}
\vskip 2cm

{\noindent{\bf Abstract}\\

A new model, which uses the frequency of individuals' annual home run
totals, is employed to predict future home run totals and records in Major
League Baseball.  Complete home run frequency data from 1903--2005 is
analyzed, resulting in annual exponential distributions whose changes can
be a used as a measure of progression in the sport and serve as a basis
against which record-setting performances can be compared.  We show that
there is an 80\% chance that Barry Bonds' current 73 home run record will
be broken in the next 10 years, despite the longevity of previous records
held by baseball legends Babe Ruth and Roger Marris.

}
\pagebreak

Previous attempts to predict athletic progressions have focused on trends
in either world records \cite{ref1,ref2} or the best annual performance
\cite{ref3}.  Our new model looks at all of the athletes and is loosely
inspired by the Gutenberg-Richter relationship for earthquake
distributions \cite{ref4}, which relates the likelihood of large, rare
events to the frequency of smaller ones.  By analyzing the entire
population, we can make predictions as to what the top performances,
interpreted as ``large events,'' will be.  We have successfully applied this
model to a variety of sports exemplifying individual achievement,
including track and field, weight lifting, and baseball \cite{ref5},
observing that the distribution of such performances is exponential.\\

In this study we examined annual home run totals for players in Major
League Baseball between 1903 and 2005 \cite{bbdata}.  For each year, we
created frequency distributions based on how many players hit a given
number of homeruns in that year (fig. 1a-c).  These distributions show
that not only are there more players hitting few homeruns, which serve as
the ``small'' events, but also that there is a relationship between this
portion of the distribution and the players with the greatest number of
homeruns.  The exponential fits shown in figure 1a-c are determined by the
95\% of players with the lowest home run totals.\\

The annual and all-time rarity of a performance is assessed by a player's
deviation from a given year's exponential distribution.  For example, the
distribution of the lowest home run totals implies that one-third of a
player would hit 51 home runs in 2005 (fig. 1c).  If we were to have three
times as many players, or simply repeat the same season three times, we
would expect to have one player hitting 51 home runs.  Andruw Jones' 2005
season can thus be viewed as a once in three-year performance, while Barry
Bonds' current record is a one in 10-year event (fig. 1b).  By
comparison, Babe Ruth's 60 home runs in 1927 is an outstanding one in
10,000-year performance (fig. 1a); far more impressive than that of Bonds,
even though the latter hit 73 home runs.  By considering relative
performances, it becomes possible to compare players of different eras,
even though baseball has progressed, largely due to improved training.  
\\

To predict home run records, we have analyzed the exponential
distributions over the past 103 years and extrapolated the progression of
these distributions to the future.  Since 1903, the parameters of the
exponential distributions have been changing in a continuous manner with
the rate of change of each being approximately constant since 1948.  We
have assumed that these rates will continue for the near future.  The top
player of a given year often performs at a level beyond what is predicted
by the lower 95\% of players.  This exceedance, which may be due to the
additional motivation that stems from lucrative financial pay-offs and
sponsorship deals that often go to the top player, is also included in our
model.  The result of these factors suggests that the probability of
somebody hitting 74 home runs within the next five years, breaking Bonds'
current record, is greater than 50\%, and over 80\% after ten years (fig.
1d). \\

\pagebreak
\noindent{\bf Acknowledgments} \\
This project was made possible through financial support from a Rains
Research Grant at Loyola Marymount University.\\

\pagebreak

\begin{figure}[h] \begin{center} \leavevmode
\includegraphics[width=1.0\textwidth]{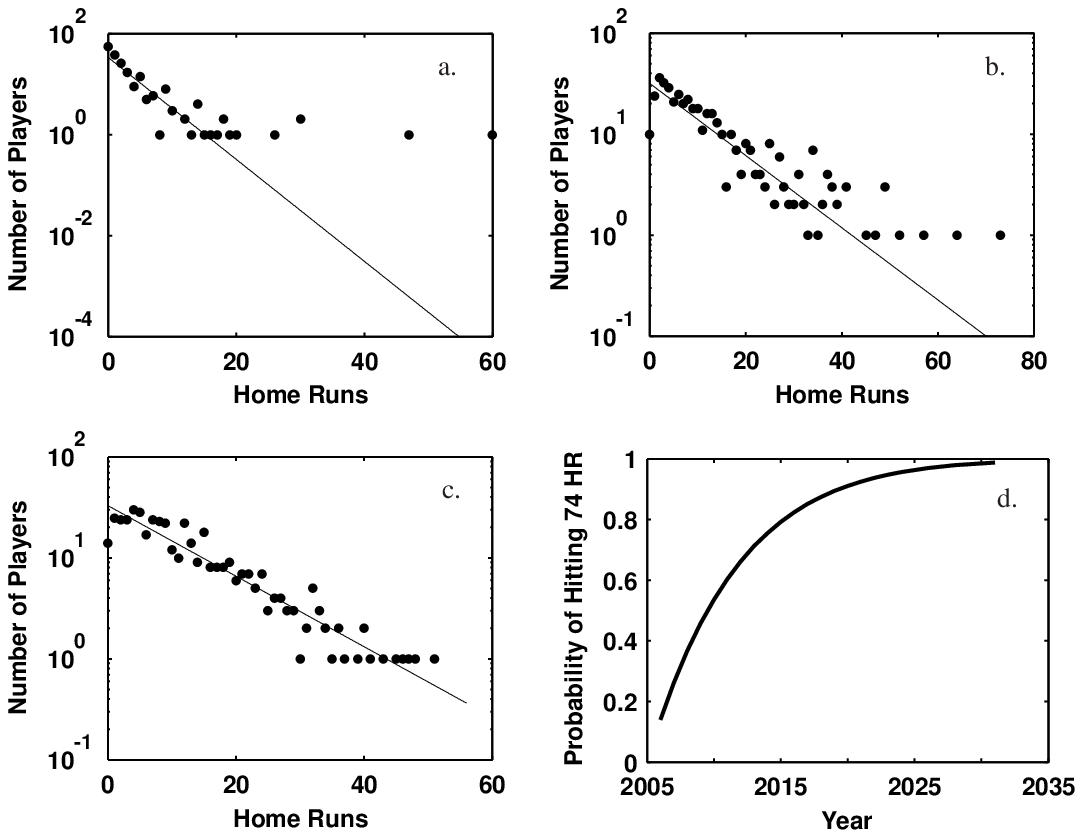} 
\caption{ Parts a, b and c show
the frequency of players hitting a number of home runs in 1927, 2001 and
2005 respectively.  To ensure that all players were playing under similar
conditions, only players with at least 100 at bats have been included.  
For all years, the frequency of the 95\% of players with the lowest home
run totals are fit with an exponential curve $be^{rN}$, where r is the
rate parameter, b is the scale and N is the number of home runs.  Part d
shows the probability that somebody will break Barry BondsÕs current home
run record before a given year. 
} 
\end{center} \end{figure}
\end{document}